# Quintuple-layer epitaxy of high-quality Bi$_2$Se$_3$ thin films for topological insulator

Guanhua Zhang, Huajun Qin, Jing Teng, Jiandong Guo, Qinlin Guo, Xi Dai, Zhong Fang and

Kehui Wu[*]

*Institute of Physics, Chinese Academy of Sciences, Beijing 100080, China*

ABSTRACT

We report the growth of atomically smooth, single crystalline Bi$_2$Se$_3$ thin films on Si(111) by using molecular beam epitaxy. Scanning tunneling microscopy, low-energy electron diffraction, X-ray photoelectron emission spectroscopy and Raman spectroscopy were used to characterize the stoichiometry and crystallinity of the film. The film grows in a self-organized quintuple-layer by quintuple-layer mode, and atomically smooth film can be obtained with the thickness down to one quintuple-layer (~1nm).

Bi$_2$Se$_3$ is a narrow gap semiconductor, which crystallizes in a rhombohedral structure belonging to the $D_{3d}^5(R\overline{3}m)$ space group. Traditionally it is an important material for thermoelectric applications [1, 2]. Very recently, it regenerates great interest by being predicted to be a 3D topological insulator (TI), a new state of quantum matter [3-10]. According to the theory, it has a large bulk energy gap and its topological surface states can be described by a single Dirac cone at the Γ point [5]. Independent ARPES study on Bi$_2$Se$_3$ did reveal such a single surface Dirac cone [6]. These results strongly suggest that Bi$_2$Se$_3$ is a fascinating topological insulator: the simple surface state makes it an ideal candidate to realize the magneto-electric effects [7], and the large bulk gap points to great potential for possible high temperature spintronics applications.

To date, all ARPES measurements on $Bi_2Q_3$ (Q=Se, Te) were done on cleaved single crystal surfaces, while for transport measurements or device applications, thin films are desired. In the literature, many works have already been done in the preparation of $Bi_2Q_3$ thin films by various techniques, such as electrodeposition [11,12], chemical bath deposition (CBD) [13], solvo thermalization [14], successive ionic layer adsorption and reaction (SILAR) [15], thermal evaporation [16], reactive evaporation [17], metal-organic chemical vapor deposition [18] and compound evaporation [19, 20], etc. However, these methods normally produce polycrystalline films aiming for thermoelectric applications, which are far from meeting the present requirements. So, growing high quality $Bi_2Se_3$ thin films is now an urgent task for the TI society.

In this paper, we report on the growth of atomically smooth, high quality, single crystalline $Bi_2Se_3$ films on Si (111) surfaces by molecular beam epitaxy (MBE) under ultrahigh vacuum (UHV) condition. More importantly, we found that the film growth exhibits an interesting self-organization phenomenon, in that the film thickness is always integer times of quintuple layers along the *c* axis. The film can be grown perfectly even with the thickness down to one quintuple layer. The high quality of the $Be_2Se_3$ film was verified by scanning tunneling microscopy (STM), low-energy electron diffraction (LEED), X-ray photoelectron emission spectroscopy (XPS) and Raman spectroscopy.

The experiments were carried out in an ultra-high vacuum system (base pressure $5\times10^{-11}$ mbar) equipped with variable temperature STM (VT-STM, Omicron), LEED, XPS, ultraviolet photoelectron emission spectroscopy (UPS) and high resolution energy loss spectroscopy (LK5000). The Si(111) subsrtrate was cut from a phosphors-doped (*n*-typed) Si wafer with the resistivity of 2 Ω•cm. Good Si(111)-7×7 surface was obtained by standard flashing the Si to about 1400 K by direct current heating. High purity Bi (99.997%) and Se(99.999%) were evaporated from Knudsen cells and the fluxes were calibrated by quartz crystal microbalance (QCM) flux monitor.

At first, we used the Si(111)-7×7 as a substrate. A room temperature co-deposition followed by post annealing was found successful to produce smooth films, as illustrated by the STM image shown in Fig. 1(a). Ordered atomic structure was also observable by STM on such a surface. However, the LEED pattern [Fig. 1(c)] shows strong background with elongated faint spots on the fringe, indicating that the film consists of flat islands, or crystalline grains, that exhibit a large rotational disorder around the c-axis, similar to the rotation of $Bi_2Te_3$ crystallites on sapphire(0001) [21]. The LEED result is verified by the STM observation of many grain boundaries in the flat film surface (Fig. 1a).

In order to further improved the film crystallinity, we tried to modify the interface to favor a better epitaxial growth. The Si(111)-7×7 reconstruction is removed by saturating the Si(111) surface with 1 monolayer Bi (here 1ML refers to the atomic density of the Si(111) plane) to form the β-√3-Bi structure [22]. In this surface the Si(111) substrate recovers to a bulk-terminated 1×1 structure underneath the 1ML Bi atoms. Such an interface is much sharper than those of films directly grown on Si(111)-7×7. Taking the √3-Bi surface as starting substrate, perfectly smooth films were obtained by co-deposition of Bi and Se at room temperature followed by annealing to 390 K. The film is atomically flat and exhibits a very sharp LEED pattern, as demonstrated in Fig. 1(b) and Fig. 1(d).

From the LEED pattern and atomically resolved STM images [see the inset of Fig. 2(a)], we can determine the lattice constant $a$ of the film to be 4.13 ± 0.10Å, quite consistent with that of $Bi_2Se_3$. More interestingly, in STM images one often sees triangular islands on flat terraces, for example the ones shown in Fig 2(a). The triangular shape reflects the hexagonal structure of the surface and it is important that all these islands exhibit identical height of 9.5 Å, as illustrated in the height profile shown in Fig. 2(c). In addition, on films with holes (not shown here), the depths of all holes is also 9.5Å. It is known that the structure of $Bi_2Se_3$ is a periodic stacking of Bi-Se quintuple layers along the $c$ axis [Fig. 2(b)], and the unit cell spans three quintuple layers with lattice constants $a \sim 4.14$ Å, $c \sim 28.64$ Å [23-24]. The observed island height and hole depth coincide with $1/3$ $c$, i.e. the height of a quintuple layer. We thus conclude that the film grows by the stacking of Bi-Se quintuple layers. This is a very important self-organization behavior that makes it easy to achieve a stoichiometric film, even with not so well controlled Bi and Se flux. In fact, the film can be grown nicely with a little bit excess of the Se flux. Moreover, by precisely controlling the coverage, we have obtained perfectly smooth films, with the thickness increases quintuple-layer by quintuple layer. The smallest thickness of the film can be tuned down to one quintuple-layer, namely ~1nm, as shown in Fig. 3.

To prove the chemical stoichiometry of our film, we measured in-situ the XPS spectra of the films. Fig. 4 shows the spectra recorded from a film of about 4 nm in thickness. The two peaks shown in Fig. 4(a) correspond to the Se $3d_{5/2}$ and $3d_{3/2}$ peaks, with binding energies of 53.5eV and 54.1eV, respectively. In Fig. 4(b), Bi $4f_{5/2}$ and $4f_{7/2}$ peaks are observed at 158.1 eV and 163.4 eV, respectively. The binding energies of the Se-3d peaks show a red shift of about 2.2eV with respect to those of pure bulk Se [25], while the binding energies of Bi-4f peaks exhibit a blue shift of about 1.1eV. The opposite shift is caused by Se-Bi bonds and charge transfer from Bi to Se. In addition, from the XPS spectra one can also estimate the surface composition by $\frac{\rho_{Se}}{\rho_{Bi}} = \frac{I_{Se}/S_{Se}}{I_{Bi}/S_{Bi}}$, where ρ is the atomic density, I is the integrated intensity of the characteristic peak and S the atomic sensitivity factor. Considering the film is very thin layers on Si substrate, surface atomic sensitivity factors are used here: $S_{Se}$ = 0.44 and $S_{Bi}$ = 2.9 [25]. We get $\rho_{Se}/\rho_{Bi}$ ~ 1.5, indicating that the films are stoichiometric.

Fig. 5 shows a typical Raman spectrum taken from a film of about 40nm in thickness. There are two characteristic peaks within the scanned frequency range, at 131.5 cm$^{-1}$ and 171.5cm$^{-1}$, which are respectively consistent with the two vibrational modes, $E_g^1$ and $A_{g1}^1$, reported for the Bi$_2$Se$_3$ single crystal [26]. Note that the $A_{g1}^2$ mode, whose frequency is 97cm$^{-1}$ is out of the measurable range (>100cm$^{-1}$) of our Raman spectroscopy.

In conclusion, we obtained high quality Bi$_2$Se$_3$ single crystalline films on Si(111) substrates by MBE. The key to improve the crystal quality is to remove the 7×7 reconstruction by using the β-√3-Bi surface as the substrate. Various techniques are used to prove that the as-grown films are stoichiometric and single crystalline. The film increases its thickness by a quintuple-layer as a unit. This work is helpful to further investigations on the physical properties, such as the 2D transport properties and thickness dependent behaviors, of topological insulators.

We thank W. R. Fang for the help in Raman measurements. This work was supported by the National Natural Science Foundation (No. 10874210) and MOST (No. 2007CB936800) of China.

**References:**

* To whom correspondence should be addressed. Email: khwu@aphy.iphy.ac.cn.

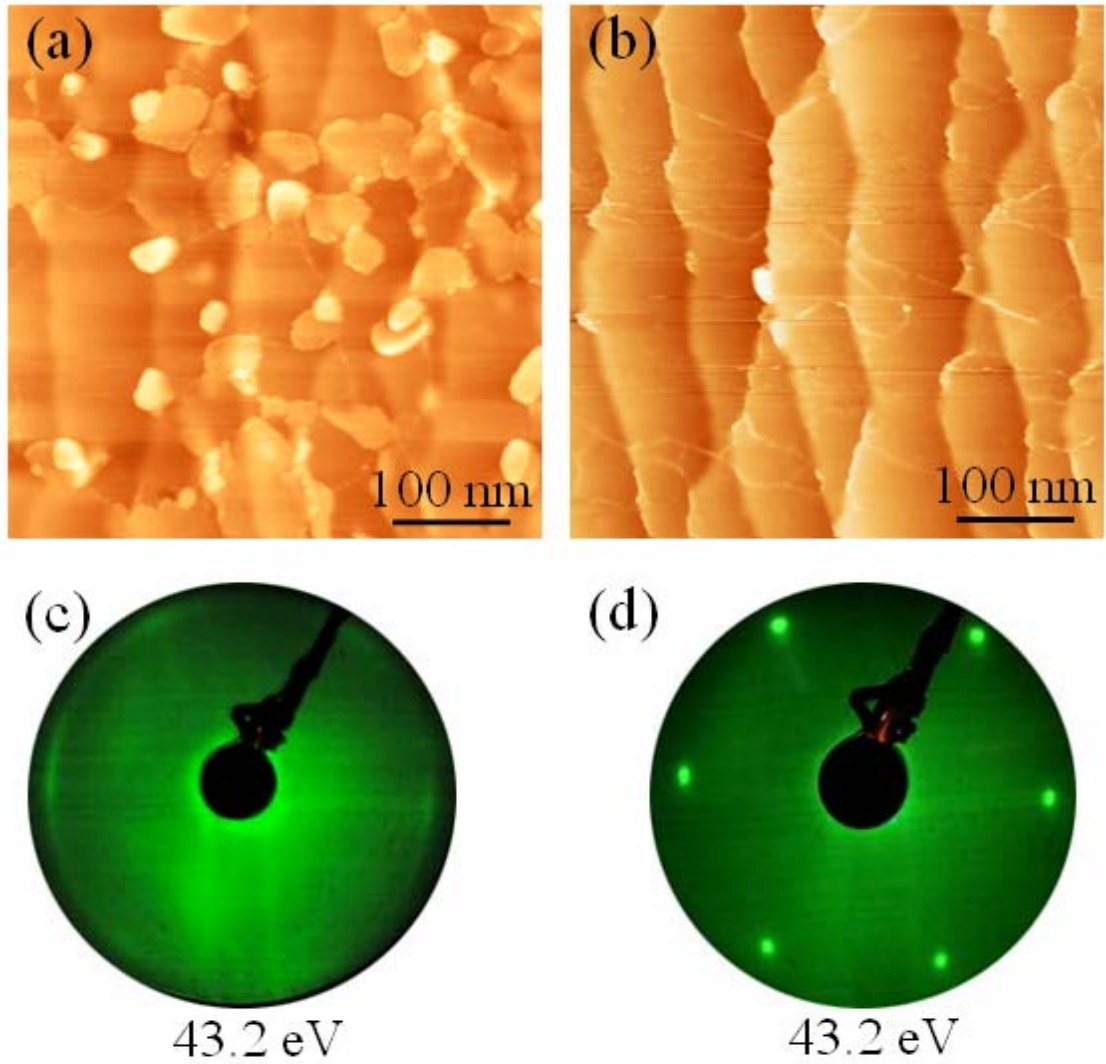

Figure 1: STM images of $Bi_2Se_3$ films (a) directly deposited on the Si-7×7 surface and (b) deposited on the β-√3-Bi substrate. (c) and (d) are LEED patterns correspond to (a) and (b) films, respectively.

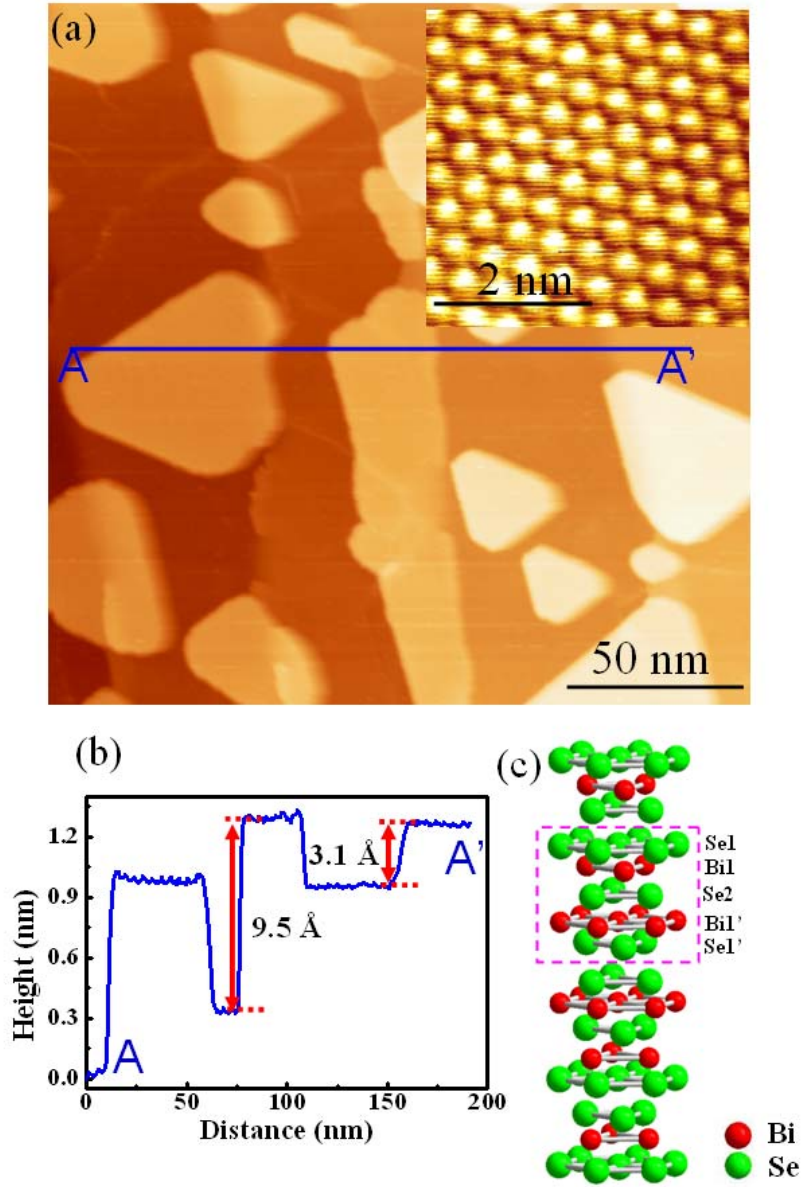

Figure 2: (a) STM image of a $Bi_2Se_3$ film of about 4 nm thick. The inset shows the atomic resolution of the surface. (b) The height profile along the AA' marked in (a) showing that the height of islands is 9.4Å. (c) Schematic crystal structure of $Bi_2Se_3$, the dashed panel shows a unit of quintuple layers.

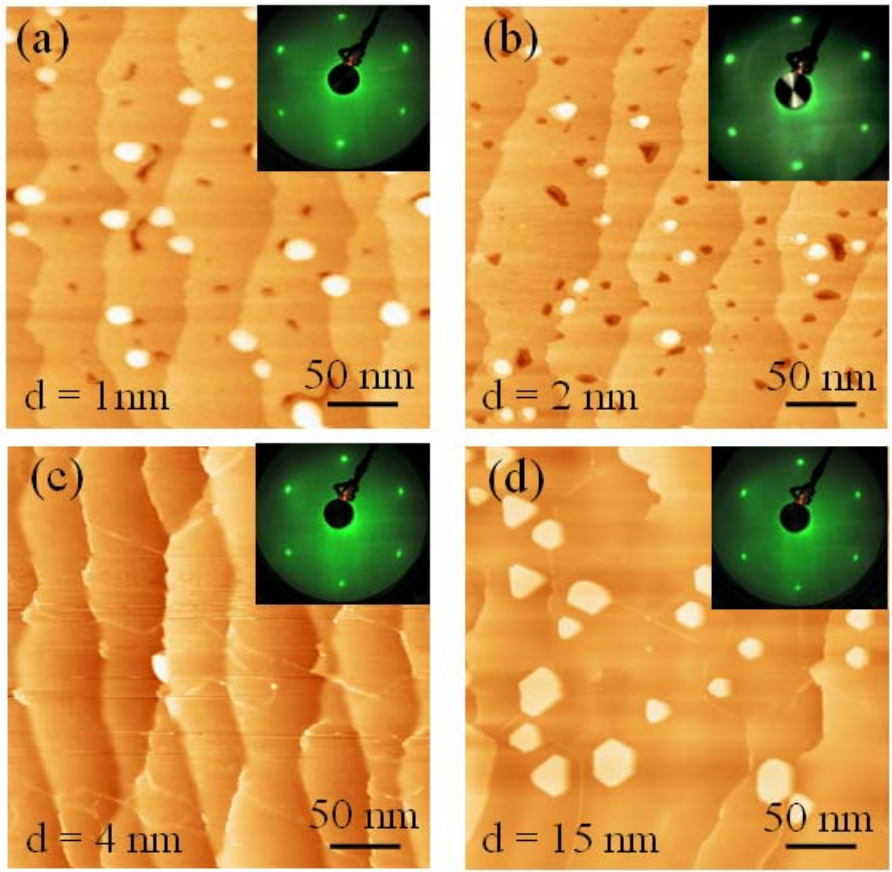

Figure 3: STM images of $Bi_2Se_3$ films grown with different thicknesses from 1 quintuple layer to 15 quintuple layers. The films are atomically smooth in large area. The steps in the images correspond to the original Si steps. Some additional small islands and holes may exist on the surface due to a slight excess or insufficient amount of $Bi_2Se_3$ deposited. These islands and holes all exhibit an identical height (depth) of 9.4Å.

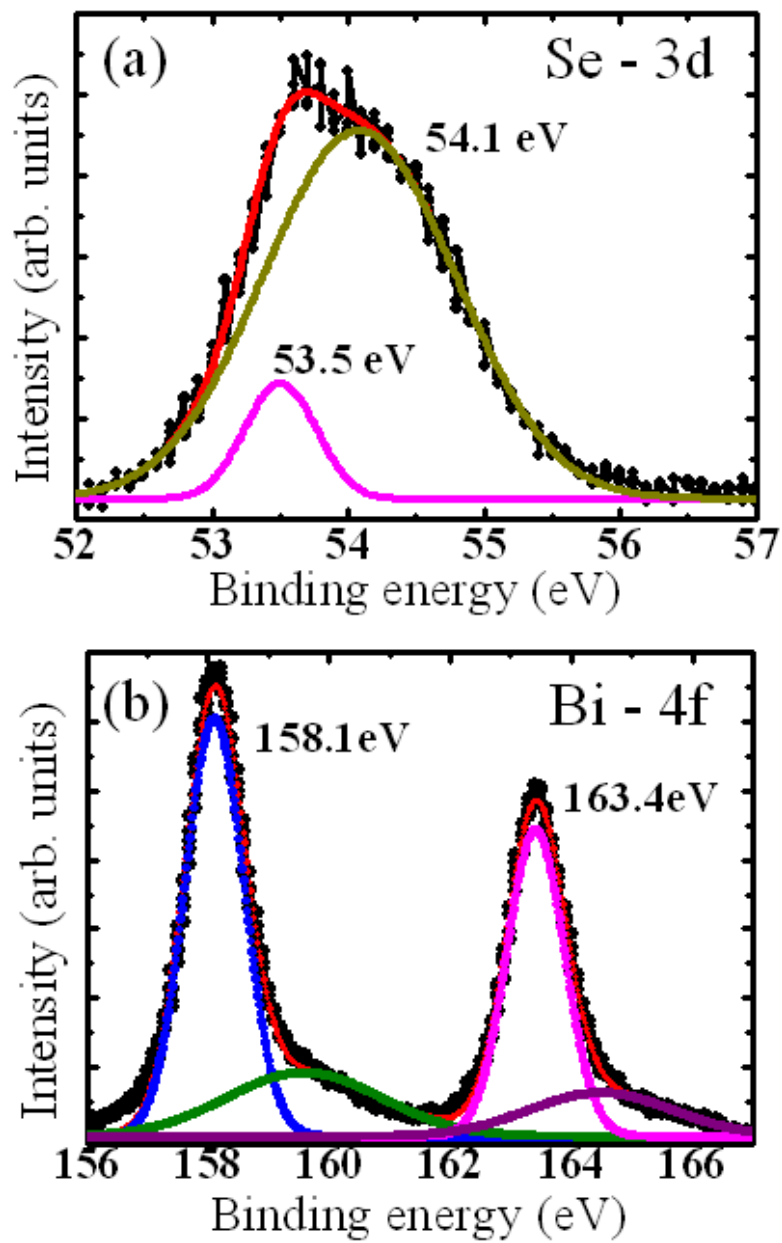

Figure 4: X-ray photoelectron emission spectra recorded from a film of about 4 nm in thickness. (a) shows the Se-3d peaks and (b) the Bi-4f peaks.

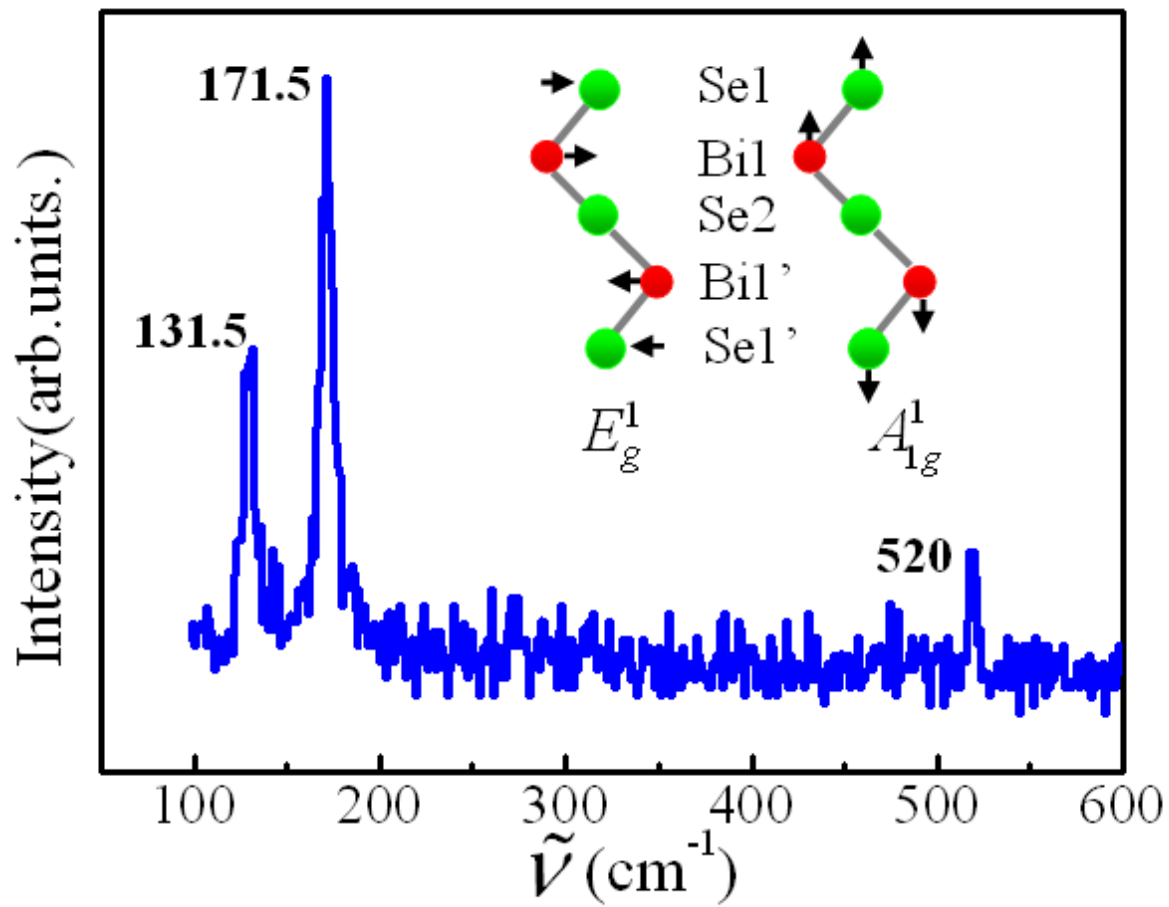

Figure 5: Raman spectrum taken from a 40 nm-thick film. The lower two peaks are assigned to the $E_g^1$ and $A_{g1}^1$ modes of $Bi_2Se_3$ crystal. The inset shows these two vibrational modes.